\documentclass[twocolumn,twoside,slac_two]{revtex4}
\usepackage{graphicx}
\usepackage{epsfig}
\usepackage{fancyhdr}
\begin{document}


\title{A Route to Understanding of the Pioneer Anomaly}

\author{$^a$Slava G. Turyshev, $^b$Michael Martin Nieto, and $^c$John D. Anderson}
 \email{turyshev@jpl.nasa.gov, mmn@lanl.gov, john.d.anderson@jpl.nasa.gov}
\affiliation{~\\
$^{a,c}$Jet Propulsion Laboratory, 
California Institute of Technology,\\
4800 Oak Grove Drive, Pasadena, CA 91109, USA\\
~\\
$^b$Theoretical Division (MS-B285), 
Los Alamos National Laboratory,\\
University of California, Los Alamos, NM 87545
}


\begin{abstract}
The Pioneer 10 and 11 spacecraft yielded the most precise navigation in deep space to date. However, while at heliocentric distance of $\sim$ 20--70 AU, the accuracies of their orbit reconstructions were limited by a small, anomalous, Doppler frequency drift.  This drift can be interpreted as a sunward constant acceleration of $a_P  = (8.74 \pm 1.33)\times 10^{-8}$ cm/s$^2$ which is now commonly known as the Pioneer anomaly.  
Here we discuss the Pioneer anomaly and present the next steps towards understanding of its origin.  They are:
1) Analysis of the entire set of existing Pioneer 10 and 11 data, obtained from launch to the last telemetry received from Pioneer 10, on 27 April 2002, when it was at a heliocentric distance of 80 AU.  This data could yield critical new information about the anomaly.  If the anomaly is confirmed, 
2) Development of an instrumental package to be operated on a deep space mission to provide an independent confirmation on the anomaly.  If further confirmed, 
3) Development of a deep-space experiment to explore the Pioneer anomaly in a dedicated mission with an accuracy for acceleration resolution at the level of $10^{-10}$ cm/s$^2$ in the extremely low frequency range.  In Appendices we give a summary of the Pioneer anomaly's characteristics, outline in more detail the steps needed to perform an analysis of the entire Pioneer data set, and also discuss the possibility of extracting some useful information from the Cassini mission cruise data.  

\end{abstract}

\maketitle

\thispagestyle{fancy}



\section{Background}
\label{intro}

The exploration of the solar system's frontiers - the region between 50-250 AU from the Sun - is a most ambitious and exciting technological challenge.  The scientific goals for possible deep-space missions are well-recognized and include studies of the gas and dust distributions, exploration of the heliopause and the space beyond, measurements of the magnetic fields and particle fluxes, studies of the Oort Cloud and Kuiper Belt Objects, encounters with distant bodies, and investigation of the dynamical background of the solar system by studying the effects of various forces that influence the trajectory of the spacecraft.  We are most interested in this last goal.  

\begin{figure*}[ht!]
 \begin{center}
\noindent    
\psfig{figure=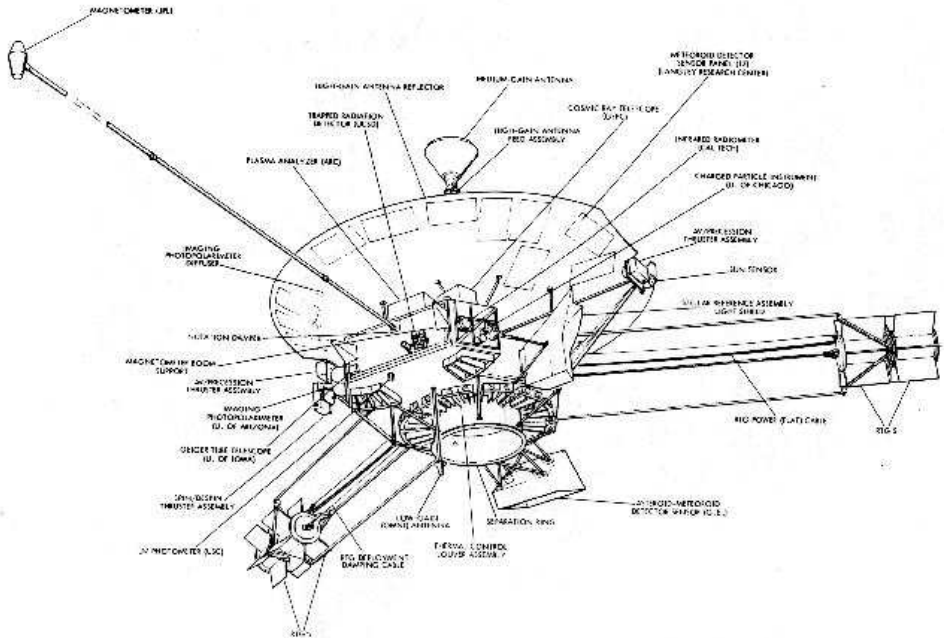,width=5in}
\end{center}
  \caption{A drawing of the Pioneer spacecraft.  
 \label{fig:trusters}}
\end{figure*}

Our interest comes from the experience working with the Pioneer 10 and 11 spacecraft - the two most precisely navigated deep-space vehicles to date.  
(See Figure \ref{fig:trusters}.)  These spacecraft had exceptional acceleration sensitivity.  However, as indicated by their radio-metric data received from heliocentric distances of 20-70 AU, the accuracies of their orbit reconstructions were limited by a small, anomalous, Doppler frequency drift 
\cite{pioprl,moriond,pioprd}.  This blue-shifted drift is uniformly changing with a rate of $(5.99 \pm 0.01)\times 10^{-9}$ Hz/s.  It can be interpreted as a sunward constant acceleration of $a_P  = (8.74 \pm 1.33)\times   10^{-8}$ cm/s$^2$ \cite{pioprd}.  This interpretation has become known as the Pioneer anomaly.  

The nature of this anomaly remains a mystery, with possible explanations ranging from normal on-board systematics to exotic gravity extensions on solar system scales. 
Although the most obvious cause would be that there is a systematic origin to the effect, the limited data analyzed does not unambiguously support any of the suggested mechanisms \cite{pioprd,old4}. 
The inability either to explain the anomaly or to test it with other spacecraft has contributed to a growing discussion about its origin.  (See the discussions in 
\cite{cospar_04,pio-mission,mex}.) 

In this paper we describe the natural steps that one would have to make in order to further understand the Pioneer anomaly.  These steps include analysis of the entire set of existing Pioneer 10 and 11 data, a small instrumental package on a large deep-space mission, and even a dedicated mission to deep space to test for the anomaly. 

The paper is organized as follows: In Section \ref{mission} we discuss the Pioneer missions and the detected anomaly. Section \ref{edata} will be devoted to a discussion of the use of data which has not been precisely analyzed, in order to further our understanding. In Section \ref{missions}  we propose new experimental tests, if the just described analysis indicates a need for them, and in Section \ref{conclude} we present our conclusion.


\section{The Pioneer Missions and the Anomaly}
\label{mission}

\begin{figure*}[ht!]
\centering  
\epsfig{figure=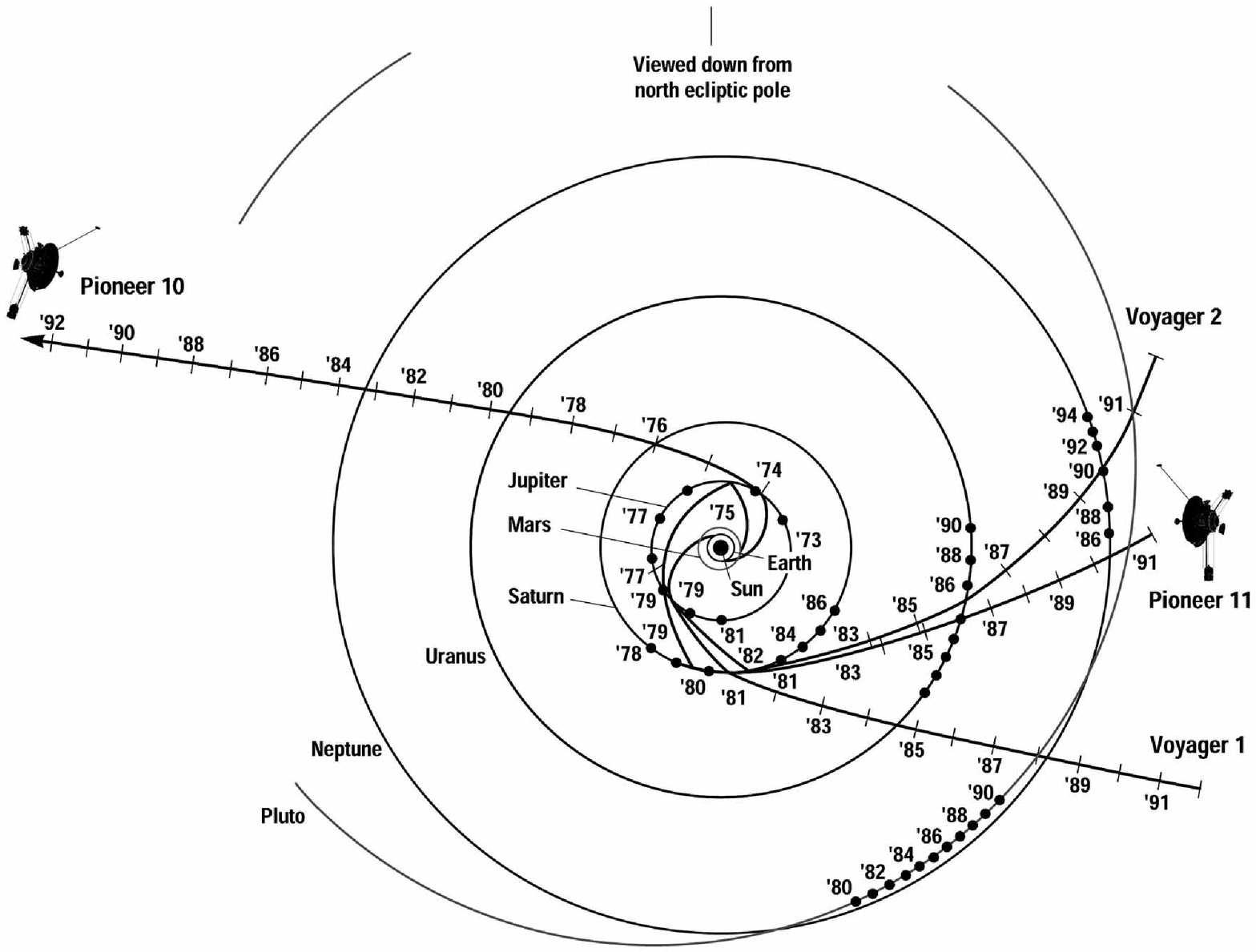,width=4.85in}
\caption
{\small
{Ecliptic pole view of Pioneer 10, Pioneer 11, and Voyager trajectories.  
Pioneer 11 is traveling approximately in the direction of the 
Sun's orbital motion about the galactic center.  The galactic center 
is approximately in the direction of the top of the figure.}} 
     \label{fig:pioneer_path}
\end{figure*}

The Pioneer 10/11 missions, launched on 2 March 1972 (Pioneer
10) and 5 April 1973 (Pioneer 11), respectively, were the first spacecraft 
to explore the outer solar system \cite{old1,bled}. After Jupiter and 
(for Pioneer 11) Saturn encounters, the two spacecraft followed escape
hyperbolic orbits near the plane of the ecliptic to opposite
sides of the solar system. 
(See Figure \ref{fig:pioneer_path}.)
Pioneer 10 eventually became the
first man-made object to leave the solar system.   Useful data was recorded even past the end of the official Pioneer 10 mission in 2001.  The last telemetry was obtained on 27 April 2002 when the craft was 80 AU from the Sun. (The last signal from the spacecraft was received on 23 January 2003.)    
	
By 1980, when Pioneer 10 passed a distance of  $\sim$ 20 AU from the Sun, the acceleration contribution from solar-radiation pressure on the craft (directed away from the Sun) had decreased to less than $4
\times 10^{-8}$ cm/s$^2$.   This meant that small effects could unambiguously be determined from the data, and the anomalous acceleration began to be seen.  
A serious study of the anomaly began in 1994, using data starting in 1987.0. By then the  external systematics (like solar-radiation pressure) were limited and the existence 
of the anomaly could unambiguously be extracted from the data.  

The initial results of the study were reported in 1998 \cite{pioprl} and
a detailed analysis appeared in 2002 \cite{pioprd}.  For this final
analysis the existing Pioneer 10/11 Doppler data from 1987.0
to 1998.5 was used \cite{pioprd}. Realizing the potential significance of the result, all {\it known} sources of a possible systematic origin for the detected anomaly were specifically addressed.  However,  even after all {\it known} systematics were accounted 
for, the conclusion remained that there was an anomalous acceleration signal of 
$a_P=(8.74 \pm 1.33) \times 10^{-8}$  cm/s$^2$ in the direction towards the Sun.   
This anomaly was a constant with respect to both time and
distance, for distances between about 20 to 70 AU from the Sun. 
(See Appendix \ref{A} for more information on the anomaly's properties.)


We emphasize {\it known} because one might naturally expect that there 
is a systematic origin to the effect, perhaps generated by the
spacecraft themselves from excessive heat or propulsion gas
leaks.  But neither we nor others with spacecraft or navigational
expertise have been able to find a convincing explanation for such a
mechanism \cite{pioprl}-\cite{old4}. 

Attempts to verify the anomaly using other spacecraft proved 
disappointing. This is because the Voyager, Galileo, Ulysses, and
Cassini  spacecraft navigation data all have their own individual difficulties for use in an independent test of the anomaly. (But see the caveat below for Cassini.)
In addition, many of the deep space missions that are currently being considered either may not provide the needed navigational accuracy and trajectory stability of under $10^{-8}$ cm/c$^2$ ((e.g., Pluto Express, but see New Horizons \cite{new-h} in Section \ref{missions}) or else they will have significant on-board systematics that mask the anomaly (e.g., JIMO -- Jupiter Icy Moons Orbiter). 

To enable a clean test of the anomaly there is also a requirement 
to have an escape hyperbolic trajectory. This makes a number of other missions (i.e., LISA -- the Laser Interferometric Space Antenna, STEP -- Space Test of Equivalence Principle,  LISA Pathfinder, etc.) less able to directly test the anomalous acceleration.  Although these missions all have excellent scientific goals and technologies, nevertheless, because of their orbits they will be in a less advantageous position to conduct a precise test of the detected anomaly. 

A number of alternative ground-based verifications of the anomaly have also been considered; for example, using Very Long Baseline Interferometry (VLBI) astrometric observations.  However, the trajectories of craft like the Pioneers, with small proper motions in the sky, making it presently impossible to accurately isolate an anomalous sunward acceleration.

To summarize, the origin of this anomaly remains unclear. 

Therefore, we advocate a program to study the Pioneer anomaly which has three phases:
\begin{itemize}
\item[i)] Analysis of the entire set of existing Pioneer 10 and 11 data, obtained from launch to the last useful communication received from Pioneer 10 in April 2002.  This data could yield critical new information about the anomaly.  If the anomaly is confirmed, 
\item[ii)] Development of an instrumental package that to be carried on another deep space mission to provide an independent confirmation for the anomaly.  If further confirmed, 
\item[iii)] Development of a deep-space experiment to explore
the Pioneer anomaly in a dedicated mission with an accuracy 
for acceleration resolution at the level of $10^{-10}$ cm/s$^2$ in the extremely low frequency  range.
\end{itemize}

In the following Sections we address the work required in these phases in more detail.


\begin{figure}[h]
 \begin{center}
\noindent 
\psfig{figure=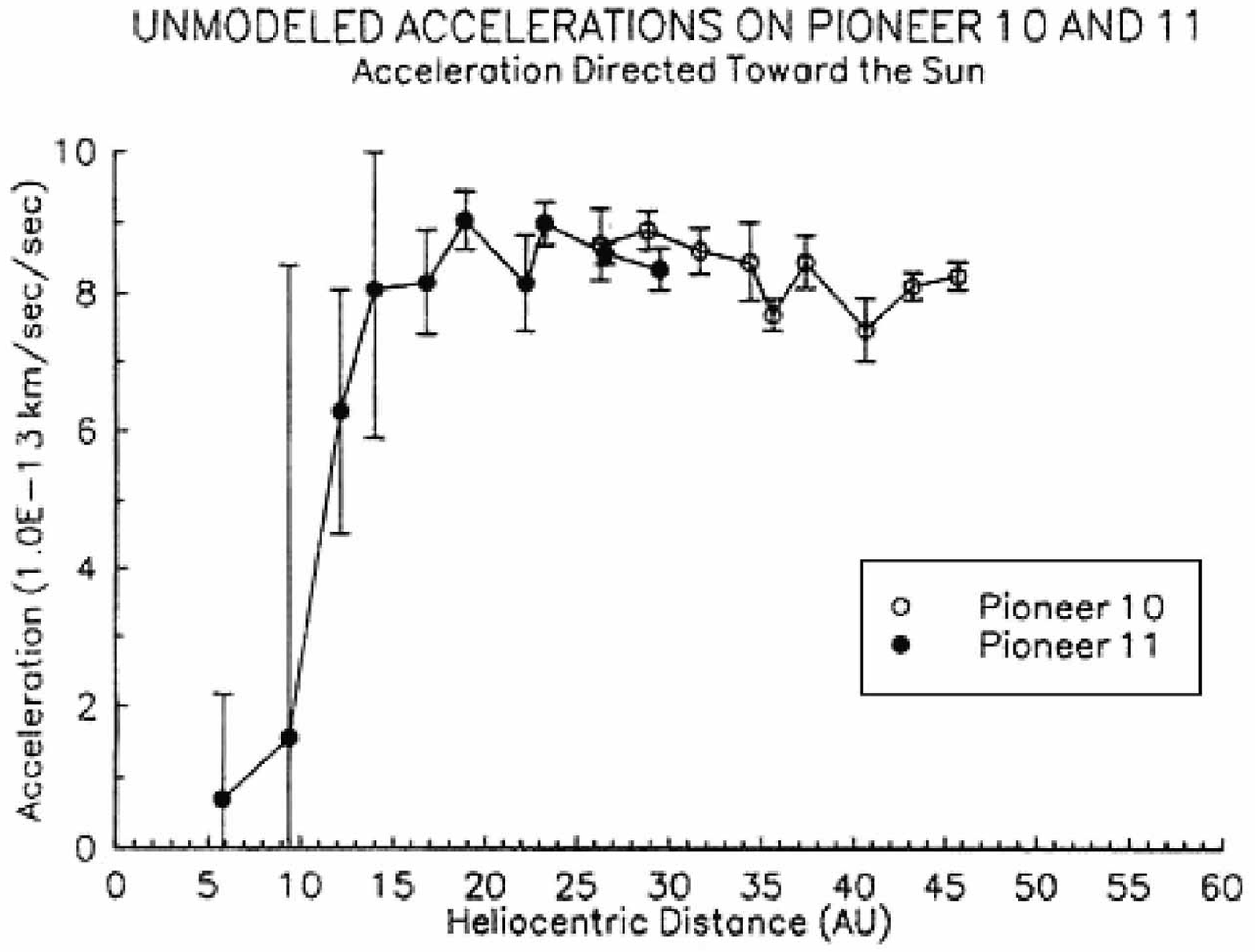,width=3.25in}
\end{center}
\vskip -10pt
  \caption{An ODP plot of the early unmodeled 
accelerations of Pioneer 10 and Pioneer 11, from about 
1981 to 1989 and 1977 to 1989, respectively.
\label{fig:correlation}}
\end{figure} 



\section{Analysis of the Entire Set of Existing Pioneer 10 and 11 Data}
\label{edata}

The early Pioneer 10 and 11 data (before 1987) were never analyzed in detail, especially with a regard for systematics.  However, by about 1980 the Doppler navigational data had began to indicate the presence of an anomaly. At first this was considered to be only an interesting navigational curiosity.  But even so. few-month samples of the data were  periodically examined by different analysts.  By 1992 an interesting string of data-points had been obtained.  They were gathered in a JPL memorandum \cite{JPLmemo}, 
and are shown in Figure \ref{fig:correlation}. 
(More details on this issue are in \cite{pioprl,pioprd}.)  

For Pioneer 10 an approximately constant anomalous acceleration seems to exist in the data as close in as 27 AU from the Sun.  Pioneer 11, beginning just after Jupiter flyby, finds  a small value for the anomaly during the Jupiter-Saturn cruise phase in the interior of solar system.  But right at Saturn encounter, when the craft passed into an hyperbolic escape orbit,  there was a fast increase in the anomaly where-after it settled into the canonical value. 

The navigation of the Pioneer spacecraft relied on an S-band radio-Doppler observable, which is not very accurate for the purposes of a 3-dimensional orbit reconstruction.  Even so, the Pioneer data from both craft when they were at shorter heliocentric distances, in to about 15 AU, indicate the presence of an anomaly in an approximate sunward direction.  

\begin{figure*}[t!]
 \begin{center}
\noindent   
\psfig{figure=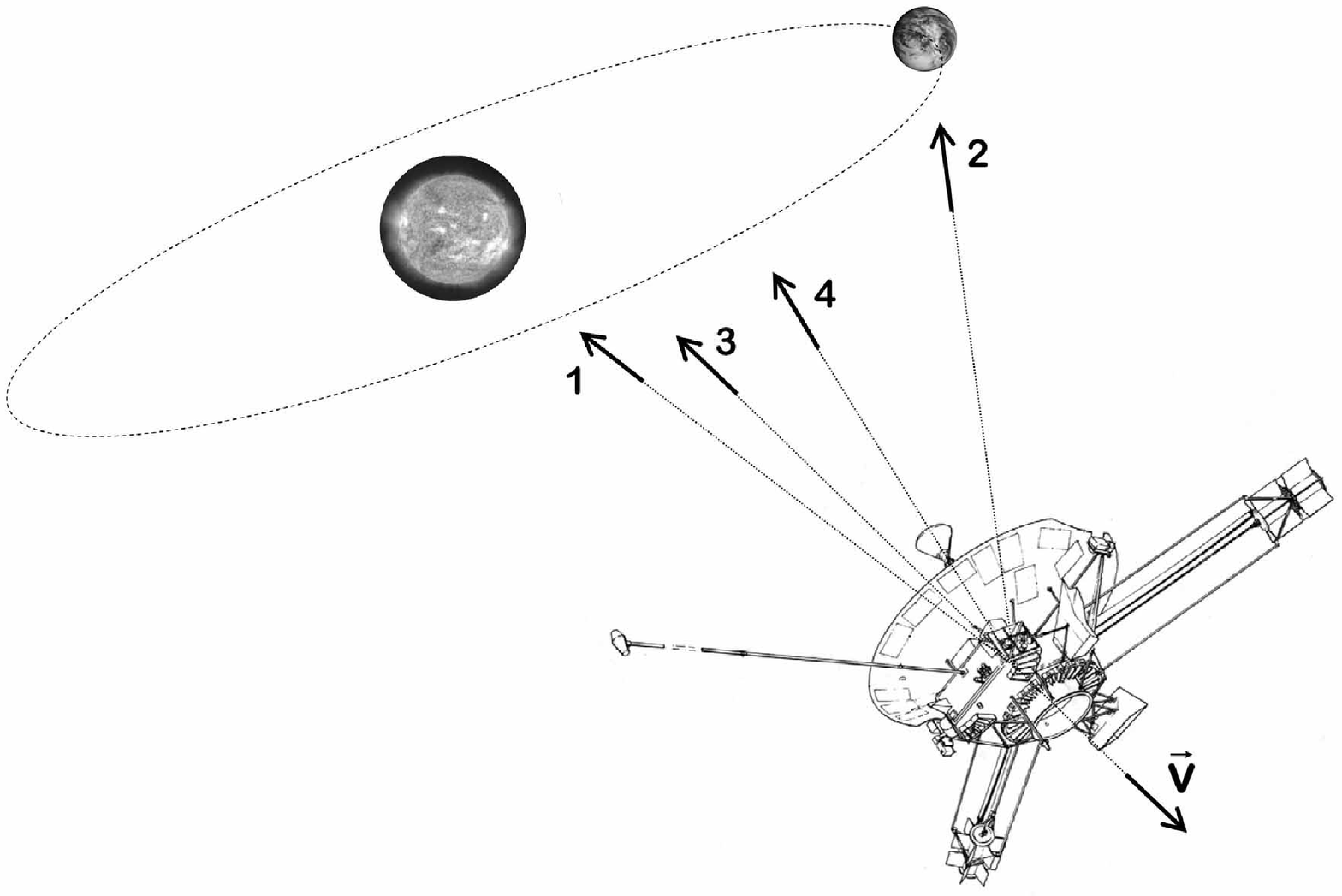,width=4.45in}
\end{center}
\vskip -10pt 
  \caption{Four possible directions for the anomalous acceleration
acting on the Pioneer spacecraft:  (1) towards the Sun, (2)
towards the Earth, (3) along the direction of motion of the craft,
or (4) along the spin axis.
 \label{fig:direction}}
%

 \begin{center}
\noindent   \vskip 1pt 
\psfig{figure=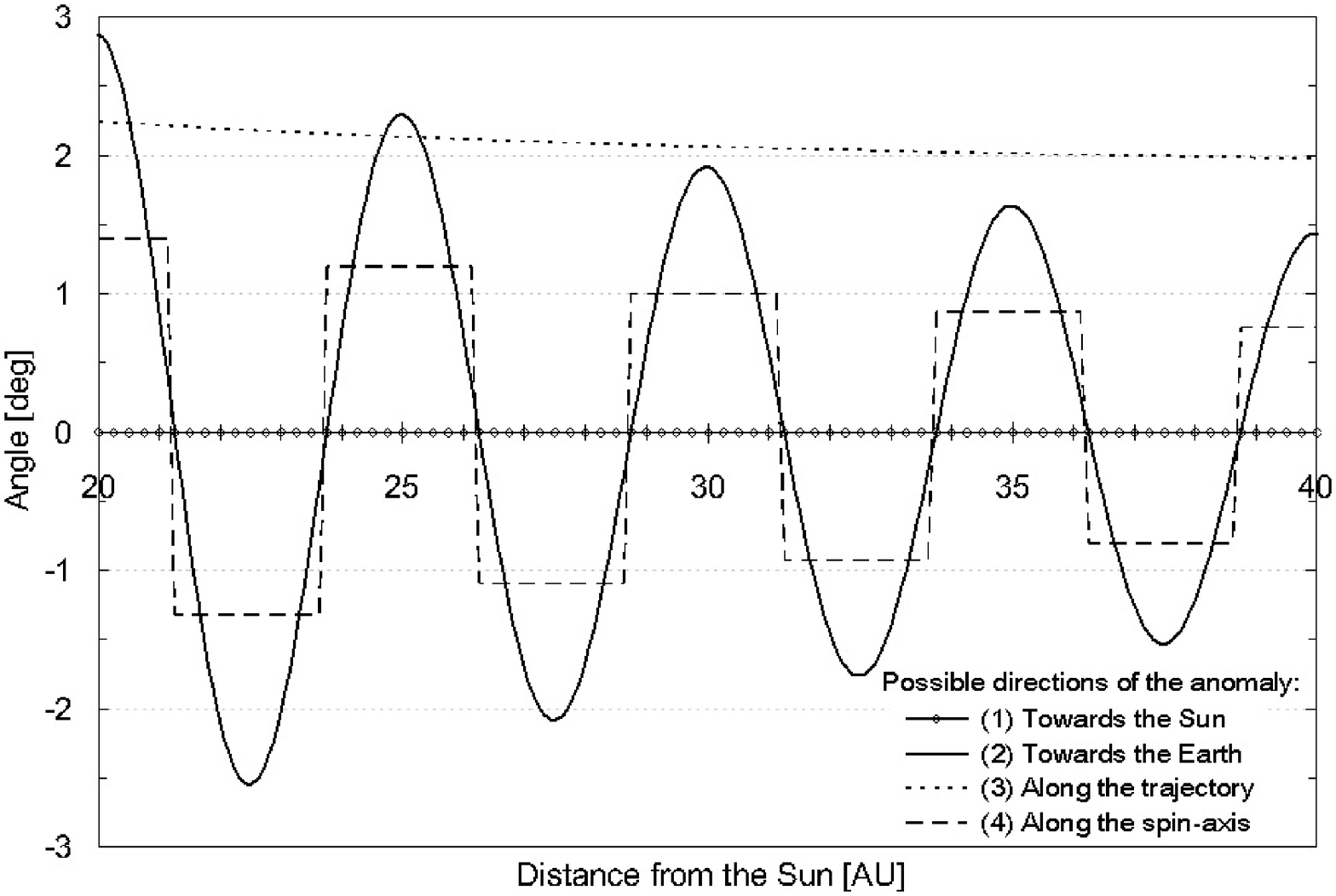,width=4.45in}
\end{center}
\vskip -15pt 
  \caption{The signatures for four possible directions of the
anomalous acceleration acting on a deep-space craft. The
signatures are distinctively different and are detectable with
precise enough data and analysis. 
 \label{fig:angles}}
\end{figure*}


With the radiation pattern of the Pioneer antennae and the lack of precise 3-D navigation, the determination of the exact direction of the anomaly was a difficult task \cite{pioprd}. While in deep space, for standard antennae without good 3-D navigation,  the directions
\begin{itemize}
\item[(1)] towards the Sun,  
\item[(2)] towards the Earth,  
\item[(3)] along the direction of motion of the craft, or  
\item[(4)] along the spin axis, 
\end{itemize}
are all observationally synonymous (see Figure \ref{fig:direction}). 

At 20 AU these directions are of order 3 degrees apart (the maximum angle subtended by the Sun and the Earth (even more depending on the hyperbolic escape velocity vector).  In Figure  \ref{fig:angles} we show the angles at which these forces would act for a hyperbolic trajectory in the ecliptic, between 20 and 40 AU \cite{pio-mission}. 

Shown in Figure \ref{fig:angles} are the following curves:

\begin{enumerate}
\item[(1)] At zero degrees, this is the reference curve indicating constant direction towards the Sun. Other angles are in reference to this.  

Starting to the right in the plane for definiteness, 

\item[(2)] shows the angle towards the Earth is a cosine curve which is modified by an $1/r$ envelope as the craft moves further out.  

\item[(3)] shows the angle from the Sun to the trajectory line.  Finally, 

\item[(4)] shows the direction along the spin axis is a
series of decreasing step functions.  This indicates two maneuvers per
year to place the antenna direction between the maximum Earth
direction and the null Sun direction, performed as the Earth passes
from one side of the Sun to the other. 
\end{enumerate}

At distances further than 40 AU, both the Sun and the Earth's orbit were within the 
$3^\circ$ of the antenna radiation pattern (set by 10 dbm range in the antenna gain), thus limiting accuracy in directional reconstruction. Therefore, analysis of the earlier data would be critical in helping to establish a precise 3-dimensional, time history of the effect.  

Looking more closely, one can understand that precise 3-dimensional 
navigation alone may give evidence to help distinguish among
the directions of interest, especially in the interior of the solar system.  

\begin{enumerate}
\item[(1)] If the anomaly is directed towards the Sun,  the aim would be to use the data to establish such a direction with sufficiently high accuracy to distinguish from the other curves.   
 
\item[(2)] If the anomaly is directed towards the Earth, the current accuracy of the Earth's ephemerides will be a key to determine this fact. Furthermore, in this case one would clearly see a dumped
sinusoidal signal that is characteristic to this situation (see above and
Figure \ref{fig:angles}).  

\item[(3)] An almost-linear angular change approaching the direction towards the Sun (also highly correlated with the hyperbolic trajectory) would indicate a trajectory-related source for the anomaly. This situation will be even more pronounced after a planetary fly-by (Note, there were three fly-byes for the Pioneers, one for Pioneer 10 and two for Pioneer 11.). In the case of a fly-by, a sudden change in the anomaly's direction will strongly suggest a trajectory-related
source for the anomaly. Finally, 

\item[(4)] A step-function-like behavior of the anomaly, strongly correlated with the maneuver history, would clearly support any anomaly directed along the spin-axis.  
\end{enumerate}

Standard navigation methods may enable one to discriminate among these four different directions of the anomaly with a sufficiently high accuracy in the interior of the solar system.  Note that because of the large external systematics, such a discrimination might not be a straightforward task.  An additional complication comes from the fact that close in to the Sun  re-orientation maneuvers were performed much more frequently than twice a year. (At times  maneuvers were performed every one and one-half months.)  But, if successfully obtained, these directions would indicate very different origins of the anomaly, corresponding to the following four possibilities: 

\begin{enumerate}
\item[(1)] new dynamical physics originating from the Sun,  

\item[(2)] anomalous behavior of frequency standards, 

\item[(3)] a drag or inertial effect, or  

\item[(4)] an on-board systematic. 
\end{enumerate}

The same investigation could study the temporal evolution of the magnitude of the anomaly. Thus, if the anomaly is due to a thermal inventory available on the vehicles, one would expect the decay of its magnitude in the manner correlated with the Plutonium decay with a half-life time of 87.74 years. The analyses of 11.5 years of data \cite{pioprd,old4} found no support for a thermal mechanism.  However, the available 30-year interval of useful data might be able to demonstrate the effect of a $\sim$21\% reduction in the fuel inventory.  This behavior, if found, would be the strongest support for a thermal origin of the anomaly. 

A critical analysis of the early data should be the first goal to understanding the anomaly. It would be relatively straight-forward and should be done first. 
In Appendix \ref{B} we describe the details of such an analysis. (In Appendix \ref{C} we observe how some useful information on the anomaly might be obtained from the 
Cassini Jupiter/Saturn cruise data.)

However, should all analysis fail to identify a systematic origin for the effect, one would then turn to ways to test for the anomaly in space.


\section{Possible Missions}
\label{missions}

\subsection{A Pioneer Instrument as Part of Another Major Mission to Deep Space}

The primary goal here would be to provide an independent experimental confirmation of the anomaly.  One can conceive of an instrument placed on a major mission to deep space.  The instrument would need to be able to compensate for systematic effects to an accuracy below the level of $10^{-8}$ cm/s$^2$.  Another concept would be a simple autonomous probe that could be jettisoned from the main vehicle, such as InterStellar Probe, presumably further out than at least the orbit of Jupiter or Saturn. The probe would then be navigated from the ground yielding a navigational accuracy below the level of $10^{-8}$ cm/s$^2$.  The data collected could provide an independent experimental verification of the anomaly's existence.

Another, more immediate possibility may be offered by NASA's New Horizons mission \cite{new-h}, which is due to launch in 2006 towards Pluto.  If enough telemetric data can be sent during cruise, and the craft is under 3-axis stabilization, a test might be feasible.

Assuming again that the Pioneer anomaly is verified, one would then consider: 


\subsection{An Experiment to Explore the Pioneer Anomaly with a Dedicated Mission}

The goal here would be to explore the anomaly at the $10^{-10}$ cm/s$^2$ level in the near DC frequency range and, in so doing, develop technologies critical for future deep-space navigation and attitude control. 

A viable concept would utilize a spacecraft pair capable of flying in a flexible formation.  The main craft would have a precision star-tracker and an accelerometer and would be capable of precise navigation, with disturbances, to a level less than 
$\sim10^{-9}$ cm/s$^2$ in the low-frequency acceleration regime.  Mounted on the front would be a disk-shaped probe with laser cornercubes embedded.  Once the configuration is on its solar system escape trajectory, will undergo no further navigation maneuvers, and is at a heliocentric distance of $\sim5-20$ AU, the co-rotating disk would be released from the primary craft.  (This concept is essentially a version of a disturbance-compensation system with a test mass being outside of the spacecraft.)  The probe will be passively laser-ranged from the primary craft with the latter having enough delta-V to maneuver with respect to the probe, if needed.  The distance from the Earth to the primary would be determined with either standard radio-metric methods operating at Ka-band or with optical communication.  Note that any dynamical noise at the primary would be a common mode contribution to the Earth-primary and primary-probe distances. This design satisfies the primary objective, which would be accomplished by the two-staged accurate navigation of the probe with sensitivity down to the $10^{-10}$ cm/s$^2$ level.  

Since the four possible anomaly directions all have
entirely different characters,  
it is clear that the use of an antenna with a highly pointed radiation pattern and star pointing sensors, will create an  even better
conditions for resolving the true direction of the anomaly than does   
the use of standard navigation techniques alone. On a spacecraft with these additional capabilities, all on-board systematics will become a common mode factor contributing to all the attitude sensors and antennas. The combination of all the attitude measurements will enable one to clearly separate the effects of the on-board systematics 
referenced to the direction towards the Sun.


\section{CONCLUSION}
\label{conclude}

The existence of the Pioneer anomaly is no longer in doubt.  Further, after much understandable hesitancy, a steadily growing part of the community has concluded that the anomaly should be understood.  Our program presents an ordered approach to doing this.  The results would be win-win; improved navigational protocols for deep space at the least, exciting new physics at the best.  Finally, a strong international collaboration would be an additional outcome of the proposed program of understanding the Pioneer anomaly.

Our first goal is to explore the Pioneer anomaly by conducting analysis of the entire set of available set of Pioneer 10 and 11 data.  The longest set includes data for Pioneer 10 from its launch in March 1972 to the last telemetric data received in March 2002.  This data is available at JPL and could yield critical new information about the anomaly. 

Simultaneously, however, we are already thinking about both an instrument aboard a larger spacecraft and a dedicated mission.


\bigskip 
\begin{acknowledgments}
The work of SGT and JDA  was carried out at the Jet Propulsion Laboratory, California Institute of Technology, under a contract with the National Aeronautics and Space Administration.  MMN acknowledges support by the U.S. Department of Energy.
\end{acknowledgments}


\appendix
\section{Summary of the Pioneer Anomaly's Properties
\label{A}}

Here we review our current knowledge of the Pioneer anomaly. 

As discussed above, the analysis of the Pioneer 10 and 11 data \cite{pioprd} demonstrated the presence of an anomalous, Doppler frequency blue-shift drift, uniformly changing with a rate of \cite{nu_dot}
\begin{equation}
\dot{\nu} \sim (5.99 \pm 0.01)\times 10^{-9} ~~ \mathrm{Hz/s}.
\end{equation}

To understand the phenomenology of the effect, consider  ${\nu}_{\tt obs}$,  the frequency  of
the re-transmitted signal observed by a DSN antennae, and 
$\nu_{\tt model}$,  the predicted frequency  of that signal.  
The observed, two-way (round-trip) anomalous effect can be expressed 
to first order in $v/c$ as  
\begin{eqnarray}
\left[\nu_{\tt obs}(t)- \nu_{\tt model}(t)\right]_{\tt DSN}
=  -2\dot{\nu}~t, 
\label{eq:delta_nu}
\end{eqnarray}
with  $\nu_{\tt model}$ being the modeled frequency change due to conventional forces accounted for in the spacecraft's motion.  (For more details see \cite{pioprd}.) This motion is outwards from the Sun and hence it produces a red shift.

After accounting for the gravitational and other large forces included in the orbital determination program \cite{pioprd} this translates to 
\begin{eqnarray}
\left[\nu_{\tt obs}(t)- \nu_{\tt model}(t)\right]_{\tt DSN}
= -\nu_{0}\frac{2a_P~t}{c}. 
\label{eq:delta_nu_syst}
\end{eqnarray}
Here $\nu_{0}$ is the reference frequency \cite{pioprd}. 

After accounting for (not modeled) systematics \cite{pioprd}, this corresponds to an anomalous acceleration of 
\begin{equation}
a_P=(8.74\pm1.33)\times 10^{-8} ~~{\rm cm/s}^2. 
\end{equation}
We have already included the sign showing that $a_P$ is inward using the DSN convention. (See Refs. [36] and [38] in \cite{pioprd} for more information.) Therefore, $a_P$ produces a slight blue shift on top of the larger red shift.

The Pioneer anomaly was studied using the following data \cite{pioprd}: 
\begin{itemize}
\item Pioneer 10: The data used was obtained between 3 January 1987 and 22 July 1998. 
This interval covers heliocentric distances  from 40 AU to 70.5 AU. This data set for Pioneer 10 had 20,055 data points obtained over the 11.5 years. 
\item Pioneer 11: The data used was obtained between 5 January 1987 to 1 October 1990. This interval covers heliocentric distances  from 22.42 AU to 31.7 AU. This data set for Pioneer 11 had 10,616 data points obtained over the 3.75 years. 
\end{itemize}
The data points were obtained using integration times ranging between 60 and 1,000 s.

By now several studies of the Pioneer Doppler navigational data have demonstrated that the anomaly is unambiguously present in the Pioneer 10 and 11 data. These studies were performed with three independent (and different!) Orbit Determination Programs (ODPs) 
\cite{pioprl,pioprd,markwardt}.  Namely: 
{}
\begin{itemize}
\item Various versions of JPL's ODP code developed in 1980-1998,
\item a version of The Aerospace Corporation's  CHASPM/POEAS code extended for deep space navigation, and finally 
\item a third code written by C.~Markward \cite{markwardt}, of the Goddard Space Flight Center (GFSC).   He analyzed Pioneer 10 data obtained from the National Space Science Data Center \cite{nssdc}, for the time period 1987-1994. 
\end{itemize}

The properties of the Pioneer anomaly can be summarized as follows \cite{pioprd}:
{}
\begin{itemize}
\item Direction: Within the 10 dbm bandwidth of the Pioneer high-gain antennae, the anomaly behaves as a line-of-sight constant acceleration of the spacecraft directed toward the Sun.
\item Distance: It is unclear how far out the anomaly goes, but the Pioneer 10 data supports the presence of the anomaly as far out as $\sim$70~AU from the Sun. In addition, the Pioneer 11 Doppler data shows the presence of the anomalous constant frequency drift as close in as $\sim$20 AU. 
\item Constancy: Temporal and spatial variations of the anomaly's magnitude are less then 3.4\% for each particular craft.
\end{itemize}

There are other pieces of information obtained from spot analyses. 
They indicate that \cite{pioprd,JPLmemo,mex}:  
\begin{itemize}
\item The anomalous acceleration was present in the Pioneer 11 data at shorter 
distances, as far in as $\sim 10$ AU. 
\item The Pioneer 11 data also indicated that the anomaly may be much smaller at distances $<10$ AU. It appears to be amplified (or turned on) at a distance of 
$\sim 10$ AU from the Sun.  This is approximately when the craft flew by Saturn and entered an hyperbolic, escape trajectory.  
\end{itemize}


\section{Description of an Expanded Data Analysis
\label{B}}

An expanded data analysis of the Pioneer 10 and 11 Doppler data should include the entire available data set.
 
\begin{itemize}
\item Pioneer 10: The entire available data set covers mission events from the launch of the spacecraft in 2 March 1972 to the last time a Pioneer 10 contact returned telemetry data, 27 April 2002. (Pioneer's last, very weak signal was received on 23 January 2003 \cite{pio_project}.) This interval spanned heliocentric distances  from $\sim1$ AU to 80 AU. The total 30 year, Pioneer 10 data set might have $\sim$80,000 data points. 

\item Pioneer 11:  The entire available data set covers from 5 April 1973  to 1 October 1990. This interval spanned heliocentric distances from $\sim$1 AU to 31.7~AU. The total  17.5 year, Pioneer 11 data set might have $\sim$50,00 data points.
\end{itemize}

To summarize, there exits about 17.5 years of Pioneer 10 and 12.5 years of 
Pioneer 11 data that was never well studied for our purposes.  
One would first have to (re)process and (re)edit the entire data span (from 1972 to 2002) at the same time, using the same initial parameters, editing strategy, and parameter estimation and noise propagation algorithms. 

In addition, one would have to process the high rate Doppler data which previously was used very little.  This particular data can better determine a spacecraft's spin rate 
and hence improve the maneuver data file information. Also, the spin rate change 
was found to be highly correlated with a small but significant spacecraft-generated 
force, probably from gas leaks \cite{pioprd}. Therefore, one would also  
have to estimate and/or calibrate valve gas leaks and all the maneuvers. 

Since the previous analysis \cite{pioprl,pioprd}, physical models for the Earth's interior and the planetary ephemeris have greatly improved. This is due to progress in GPS- and VLBI-enabling technologies, Doppler spacecraft tracking, and new radio-science data processing algorithms. One would have to write and/or update existing orbit determination programs using these latest Earth models (adopted by the IERS) and also  using the latest planetary ephemeris. This will improve the solutions for the DSN ground station locations by two orders of magnitude (1 cm) over that of the previous analysis. Additionally, this will allow a better characterization of not only the constant part of any anomalous acceleration, but also of the annual and diurnal terms detected in the Pioneer 10 and 11 Doppler residuals \cite{moriond,pioprd}. 

One would expect that the much longer data span will greatly improve the ability to determine the source of the acceleration. In particular, 
with data from much closer to the Earth and Sun, one should be able to better determine whether the acceleration is directed towards the Earth, the Sun, or along the spacecrafts' spin axes or velocity vectors, and therefore whether it is due to a systematic or new physics.  Further, with the early data (where spacecraft re-orientation maneuvers were performed much more often than twice per year), we expect to improve the sensitivity of the solutions in the directions perpendicular to the line-of-sight by at least an order of magnitude. 

Finally, of course, the question of collimated thermal emission would have to be revisited.  The much longer data span of 30 years will help to better determine if there is any signature of an exponential decay of the on-board power source, something not seen with the 11.5 years of data. 

Therefore, the extended data set, augmented by all the ancillary spacecraft data, will help not only to precisely identify the direction of the anomaly but also will help to obtain tighter bounds on its time and distance dependence.   
This wealth of additional data available presents an exciting opportunity to learn more about the Pioneer anomaly in various regimes and, thus, help to determine the nature of the anomalous signal.


\section{THE CASSINI CRUISE DATA
\label{C}}

Recall that inside 10 AU, the old, not thoroughly analyzed data 
(see Figure \ref{fig:correlation}) seems to indicate very little anomaly as Pioneer 11 was on its bound-orbit, Jupiter/Saturn cruise.  Cassini was also on a bound-orbit, Jupiter/Saturn cruise \cite{casorbit}. Therefore, one might be able to perform a partial test of the Pioneer anomaly using Cassini cruise data, to see if it, also, finds very little anomaly on this trajectory.    

Even given that the reaction wheels (needed during good navigational data taking periods) 
do not overly affect the data and power usage as they are turned on and off, 
the biggest problems would be:  
\begin{enumerate}
\item[(1)] to disentangle the large heat systematic caused by the RTGs mounted forward on the craft, 
\item[(2)] the 750 W of electrical power that has to be dissipated, and 
\item[(3)] the changing mass of the craft.
\end{enumerate}

The total weight of the craft at launch was about 5,600 kg, of which
more than half (3,132 kg) was fuel.  The three Cassini RTGs are mounted 
directly in front of the craft.  They are of the GPHS type also used on Galileo. They generate about 4 kW of heat each or about 12 kW from all three. (They make about 250 W of electrical power each.)  From 
the spacecraft geometry one can estimate that the
fractional excess power creating an acceleration backward is   
on the order of 3,000 W.  Given that about half the mass of the fuel was gone during cruise, one can estimate that the systematic from the RTGs is on the order of 
$\sim 3 ~ a_P \sim 27 \times 10^{-8}$ cm/s$^2$.

At this meeting it was reported that good navigational data has been forthcoming during Cassini's Saturn orbit \cite{luciano}.  Since the angle of the antenna towards the Earth must be maintained, the direction of the systematic force on Cassini will be changing with respect to Saturn's gravitational force on Cassini. This should allow a calibration of the systematic to be made.  Then this calibration can be used in an attempt to precisely model the navigation during Jupiter/Saturn cruise. We encourage the Cassini Radio Science and Navigation Teams to conduct such an investigation.


\bigskip 


\begin{thebibliography}{99}
\bibitem{nu_dot} 
Depending on the particular piece of data and the fitting procedure, the exact size and formal error vary slightly. (This particular number comes from the ``experimental" result for $a_P$ determined with Pioneer 10 data as described in Section VI of Ref. \cite{pioprd} and is referenced to the downlink carrier frequency, 2.29 GHz.)  
But these differences are much less that the size of the anomaly and also significantly less than the systematic error.
\end{thebibliography}


\end{document}